\documentclass[12pt]{article}
\usepackage{graphicx}
\begin{document}
\title{Classical signal-flow in cluster-state quantum computation}
\author{Kazuto Oshima\thanks{E-mail: oshima@elc.gunma-ct.ac.jp}    \\ \\
\sl Gunma National College of Technology, Maebashi 371-8530, Japan }
\date{}
\maketitle
\begin{abstract}
We study concretely how classical signals should be processed 
in quantum cluster-state computation. Deforming corresponding quantum 
teleportation circuit, we find a simple rule of a classical signal-flow 
to obtain correct quantum computation results.
\end{abstract} 
\newpage
Cluster-state quantum computation proposed by Raussendorf and Briegel\cite{Rau} is 
a promising scheme for its simplicity. Preparing a cluster-state corresponding to
a certain problem, we can perform quantum computation only by successive quantum
measurements and feedforward of measurement outcomes. We can obtain a proper computation
result only through choosing adequate quantum measurement basis and rectifying a
resultant quantum state. These choices of the basis and rectification of the resultant
quantum state are done referring to preceding quantum measurement outcomes.
Therefore, it is inevitable to treat quantum measurement outcomes properly. As far as
the author knows, however, treatment of measurement outcomes has not been studied
significantly. The purpose in this paper is to give a concrete classical
signal-flow chart that leads to proper computation results in cluster-state quantum computation.

Cluster-state quantum computation is nothing but quantum computation with
measurements \cite{Gott,KLM}. Actually, we can associate a quantum teleportation circuit with
a cluster-state. For a quantum teleportation circuit, dragging back controlled-phase
transformations connecting pairs of junction qubits to the starting point, we obtain 
the corresponding cluster state.  We can find a classical signal-flow in cluster-state
computation by analyzing the corresponding quantum teleportation circuit.

It is well known that a set of
single-qubit unitary transformations and the two-qubits controlled-phase transformation is
universal. An arbitrary single-qubit unitary transformation can be carried out by
a sequence of three elements in the set $\{HZ_{\alpha}\}$, where $H$ is
the Hadamard transformation and $Z_{\alpha}=\cos{\alpha \over 2}I-i\sin{\alpha \over 2}Z$,
with $Z=|0 \rangle\langle 0|-|1 \rangle\langle 1|$ for quantum computational
basis states $\{|0\rangle, |1 \rangle\}$. This can be seen by noting the identity
$HZ_{\alpha} HZ_{\beta} HZ_{\gamma}= HZ_{\alpha}X_{\beta}Z_{\gamma}$, where
$X_{\beta}=\cos{\beta \over 2}I-i\sin{\beta \over 2}X$, with
$X=|0 \rangle\langle 1|+|0 \rangle\langle 1|$.

A single-qubit unitary transformation $HZ_{\alpha}$ on a quantum state
$|\psi\rangle$ can be realized by the single-qubit quantum teleportation
circuit\cite{Nielsen-Chuang}. The single-qubit teleportation circuit is
nothing but the simplest cluster-state quantum computation.  The first qubit of
the initial cluster state $|\psi\rangle-|+\rangle$, where the $-$ sign indicates
the controlled-phase transformation, is measured by the basis
$\{(HZ_{\alpha})^{\dagger}|0\rangle, (HZ_{\alpha})^{\dagger}|1\rangle\}$ to transform
the second qubit into $HZ_{\alpha}|\psi\rangle$.

The single-qubit quantum teleportation circuits can be combined into a multistage
type\cite{Nielsen} corresponds to a one-dimensional cluster state. Since a product of three successive transformations
$HZ_{\alpha_{1}} HZ_{\alpha_{2}} HZ_{\alpha_{3}}$ causes an arbitrary single-qubit unitary transformation, the number
of the sequence can be limited to three. 

Starting from an adequate cluster state, we can carry out desired quantum computation
by precise quantum measurements and proper processing of measurement outcomes.
Each controlled-phase transformation in an associated quantum teleportation circuit
corresponds to a controlled-phase transformation
between a pair of junction qubits in a cluster state.  A multistage single-qubit quantum 
teleportation circuit replaces a one-dimensional chain in a cluster state. This quantum 
teleportation circuit shows us how quantum measurement outcomes should be treated 
in cluster-state computation.

In a cluster state that we are considering, junction qubits appear in pairs connected by 
controlled-phase transformations.  Each juncton qubit connects a one-dimensional incoming chain with
a one-dimensional outgoing chain. Therefore, a pair of junction qubits forms a fundamental
block of an H-branch with two one-dimensional incoming chains and two one-dimensional
outgoing chains. Two H-branches are connected by sharing a one-dimensional chain
that is outgoing from one H-branch and incoming into the other H-branch.
Connecting H-branches successively we can form a desired cluster state.
Successive measurements of qubits cause a classical signal-flow over the cluster state.
Classical signals flow over a one-dimensional chain and branch at junctions.
Our purpose is to obtain a classical signal-flow chart for a cluster
state using the corresponding quantum teleportation circuit.

First, we study a classical signal-flow for a one-dimensional quantum chain that is
simulated by a multistage quantum teleportation circuit.
The first qubit on the left is in a state $|\psi\rangle$ and the other qubits are in
the state $|+ \rangle={1 \over \sqrt{2}}(|0\rangle+|1\rangle)$ untill the
controlled-phase transformations are carried out.
Except for the last qubit or the last few qubits, we measure each qubit by the basis
$(HZ_{\alpha_{i}})^{\dagger}|0\rangle, (HZ_{\alpha_{i}})^{\dagger}|1\rangle (i=1,2,\cdots)$
from the left.

As is well-known, to teleport a quantum state successfully, we need to operate 
the Pauli matrices $X$ and $Z$ according to measurement outcomes.
In a multistage quantum teleportation circuit we need to teleport intermediate quantum states
that are corrected by the Pauli matrices.
Using the identity $(HZ_{\alpha}\otimes I)C_{\rm phase}(Z \otimes I)
=(X \otimes I)(HZ_{\alpha}\otimes I)C_{\rm phase}$, that comes from $HZ=XH$,
the correction $Z$ in the first qubit can be changed into a correction $X$ after
the controlled-phase transformation and the $HZ_{\alpha}$ transformation. 
In the same way using the identity $(HZ_{\alpha}\otimes I)C_{\rm phase}(X \otimes I)
=(Z \otimes Z)(HZ_{-\alpha}\otimes I)C_{\rm phase}$, the correction $X$ can be postponed.
Moreover, we can see that the operation $Z \otimes Z$ can be removed by adding
a $Z$ after the controlled-$X$ operation in the second qubit.  
Thus the quantum teleportation circuit in Fig.1(a) can be transformed into the circuit
in Fig.1(b). If the measurement outcome of the $(n-1)$-th qubit is $1$,
the $n$-th qubit is measured after the $HZ_{-\alpha}$ transformation instead of the
$HZ_{\alpha}$ transformation. If the measurement outcome of the $(n-2)$-th qubit is $1$,
the additional transformation $X$ is required before the meter.

The situation is simple when the present qubit is not teleported.
If the $n$-th qubit is not measured, the $(n-1)$-th qubit measurement outcome $1$
brings the $X$ correction to the $n$-th qubit, and the $(n-2)$-th qubit measurement
outcome $1$ brings the $Z$ correction to the $n$-th qubit. Even if the $(n-1)$-th qubit
is not measured the $n$-th qubit is corrected by $Z$ from the the $(n-2)$-th qubit
measurement outcome $1$.

In anyway, it is important to note that the measurement outcome of a qubit affects
only the following two qubits. This rule holds also for two-dimensional cluster states.

Second, we consider two-dimensional cluster states. The H-branch in Fig.2(a) is
a fundamental constituent of two-dimensional cluster states. We study how the measurement
outcomes affect the qubits around the junctions. The only distinct of the
two-dimensional case from the one-dimensional case originates from the
controlled-phase transformation connecting a pair of junction qubits.
In cluster-state quantum computation these controlled-phase transformations have been done
at the outset. The corresponding quantum teleportation circuit with measurements
having been done up to the left side of the pair of junction qubits is depicted
in Fig.2(b). In Fig.2(b), two single-qubit quantum-teleportation circuits are connected
by the central controlled-phase transformation.
The controlled-phase transformation connecting the two junction
qubits can be dragged back to the left as the controlled-phase transformation 
in the quantum teleportation circuit. The only extra effect is to
add the extra $X$ transformation on the $n$-th($m$-th)
qubit due to a measurement outcome $1$ of the $(m-1)$-th($(n-1)$-th) qubit. If both of
the measurement outcomes of the $(m-1)$-th and the $(n-1)$-th qubits are $1$, an extra
harmless overall minus factor appears due to an exchange of $Z$ and $X$. In Fig.4(c)
the two extra $X$ corrections through the junctions are depicted before the meters.
Note that the $(n+1)$-th qubit, which is next to the junction qubit, is not affected by
the $m$-th qubit. If the $n$-th qubit is not measured this qubit is corrected by $X$
from the $(n-1)$-th qubit measurement outcome $1$ and by $Z$ from the $(n-2)$-th qubit
measurement outcome $1$ and by $Z$ from the $(m-1)$-th qubit measurement outcome $1$.

Our result of classical signal-flow is summarized in Table~I. Table~I shows how the
$n$-th qubit should be corrected according to the measurement outcomes of prior qubits.
This table partially applies also for the case the $n$-th qubit not being a junction qubit.
\\ \\
{\bf TABLE~I.} Corrections to the $n$-th qubit of the H-branch in Fig.2(a).
The first column indicates control qubits. The middle row has meaning only
when the $(n-1)$-th qubit is measured. The final row has meaning only when
the $n$-th qubit is a junction qubit. \\
\begin{tabular}{c c c}\hline \hline
$n$ &   Not Measured &    Measured \\ \hline 
$n-2$ &  $Z$        & \quad $X$ before the meter \\ 
$n-1$ & $X$        & \quad $Z_{\alpha} \rightarrow  Z_{-\alpha}$\\ 
$m-1$ & $Z$         & \quad $X$ before the meter  \\ \hline \hline
\end{tabular}
\\ \\

\newpage
FIGURE CAPTIONS\\
FIG~1. 
(a)The single-qubit quantum teleportation circuit with the $ZX$ correction 
in the incoming state. The vertical line means the controlled-pase transformation 
$C_{\rm phase}$. \qquad
 (b)A transformed circuit of (a). The arrows mean classical
signal-flow.\\
\\
FIG~2.(a)An H-branch.  (b),(c)Quantum teleportation circuits that simulate the H-branch.\\
\end{document}